\def\arXiv{}

\documentclass{article}
\usepackage{ijcai18}

\usepackage{times}
\usepackage[dvipsnames]{xcolor}
\usepackage{soul}
\usepackage[utf8]{inputenc}
\usepackage[small]{caption}

\ifdefined\arXiv
\else
\def\anon{}
\fi

\ifdefined\anon
\else
\fi

\usepackage{multicol}
\usepackage{multirow}

\usepackage{longtable}
\usepackage[T1]{fontenc}    %

\usepackage{url}

\usepackage{breakurl}
\usepackage[breaklinks]{hyperref}

\usepackage{booktabs}       %
\usepackage{amsfonts}       %
\usepackage{nicefrac}       %
\usepackage{microtype}      %

\usepackage{amsmath}
\usepackage{amssymb}
\usepackage{listings}

\usepackage{tikz}
\usetikzlibrary{
  shapes,calc,arrows,fit,positioning,decorations.pathmorphing,snakes,intersections,shapes.geometric,trees,cd
}

\usepackage{subcaption}
\usepackage[usenames,dvipsnames]{pstricks}
\usepackage{epsfig}
\usepackage{pst-grad} %
\usepackage{pst-plot} %
\usepackage[space]{grffile} %
\usepackage{etoolbox} %
\makeatletter %
\patchcmd\Gread@eps{\@inputcheck#1 }{\@inputcheck"#1"\relax}{}{}
\makeatother

\newcommand{\code}[1]{\lstinline!#1!}

\newcommand{\stacklabel}[1]{\stackrel{\smash{\scriptscriptstyle \mathrm{#1}}}}
\newcommand{\defeq}{\stacklabel{def}=}

\newcommand{\refsec}[1]{\S\ref{#1}}

\definecolor{mygray}{rgb}{0.5,0.5,0.5}
\lstdefinestyle{customlisp}{
  belowcaptionskip=1\baselineskip,
  breaklines=true,
  language=Lisp,
  showstringspaces=false,
  numbers=left,
  xleftmargin=2em,
  framexleftmargin=1.5em,
  numbersep=5pt,
  numberstyle=\tiny\color{mygray},
  basicstyle=\small\ttfamily,
  keywordstyle=\color{blue},
  commentstyle=\itshape\color{purple!40!black},
  stringstyle=\color{orange},
  morekeywords={def-rule, ?},
  tabsize=2
}

\usepackage{xcolor}
\usepackage{soul}
\definecolor{LightGreen}{rgb}{0.75,1.00,0.75}
\definecolor{LightOrange}{rgb}{1.00,0.90,0.50}
\definecolor{LightRed}{rgb}{1.00,0.75,0.75}
\definecolor{LightBlue}{rgb}{0.75,0.75,1.00}
\DeclareRobustCommand{\commentformat}[3]{\sethlcolor{#2}\hl{\textsf{#1: #3}}}

\newcommand{\sophia}[1]{{\commentformat{SK}{LightOrange}{#1}}}

\newcommand{\pxm}   [1]{{\commentformat{PM}{LightGreen} {#1}}}

\def\draft{}

\ifdefined\final
\undef\draft{}
\fi

\ifdefined\arXiv
\undef\draft{}
\fi

\ifdefined\draft
\else
\renewcommand{\commentformat}[3]{}
\fi

\begin{document}

\title{Latent Factor Interpretations for Collaborative Filtering}
\ifdefined\arXiv
\author{
Anupam Datta, Sophia Kovaleva, Piotr Mardziel, Shayak Sen\\
Carnegie Mellon University
}
\else
\author{
Anon Anonymous\\
Anonymous University
}
\fi
\maketitle

\begin{abstract}

  Many machine learning systems utilize latent factors as internal
  representations for making predictions. 
  Since these latent factors are largely uninterpreted, however,
  predictions made using them are opaque.
  Collaborative filtering via matrix factorization is a prime example
  of such an algorithm that uses uninterpreted latent features, and
  yet has seen widespread adoption for many recommendation tasks.

  We present Latent Factor Interpretation (LFI), a method for
  interpreting models by leveraging interpretations of latent factors
  in terms of human-understandable features. 
  The interpretation of latent factors can then replace the
  uninterpreted latent factors, resulting in a new model that
  expresses predictions in terms of interpretable features. 
  This new model can then be interpreted using recently developed
  model explanation techniques.
  In this paper we develop LFI for collaborative filtering based
  recommender systems.%

  We illustrate the use of LFI interpretations on the MovieLens
  dataset, integrating auxiliary features from IMDB and DB tropes, and
  show that latent factors can be predicted with sufficient accuracy
  for replicating the predictions of the true model.

\end{abstract}

\section{Introduction}\label{sec:intro}

Many machine learning systems utilize latent factors as internal
representations for making predictions. 
Since these latent factors are largely uninterpreted, however,
predictions made using them are opaque. 
Recommender systems that perform collaborative filtering via matrix
factorization are prime examples of such machine learning systems. 
These systems are state-of-the-art in important application domains,
including movie and social
recommendations~\cite{koren2009matrix,facebook-cf}.
However, these models are difficult to interpret because they express
user preferences and item characteristics along a set of uninterpreted
latent factors trained from a sparse set of user ratings.

We present Latent Factor Interpretation (LFI), a method for
interpreting models by leveraging expressions of latent factors in
terms of human-understandable features, and develop it in the
particular setting of collaborative filtering.
In order to interpret models that use uninterpreted latent factors, we
address three challenges.

\begin{figure}
  \includegraphics[width=\columnwidth]{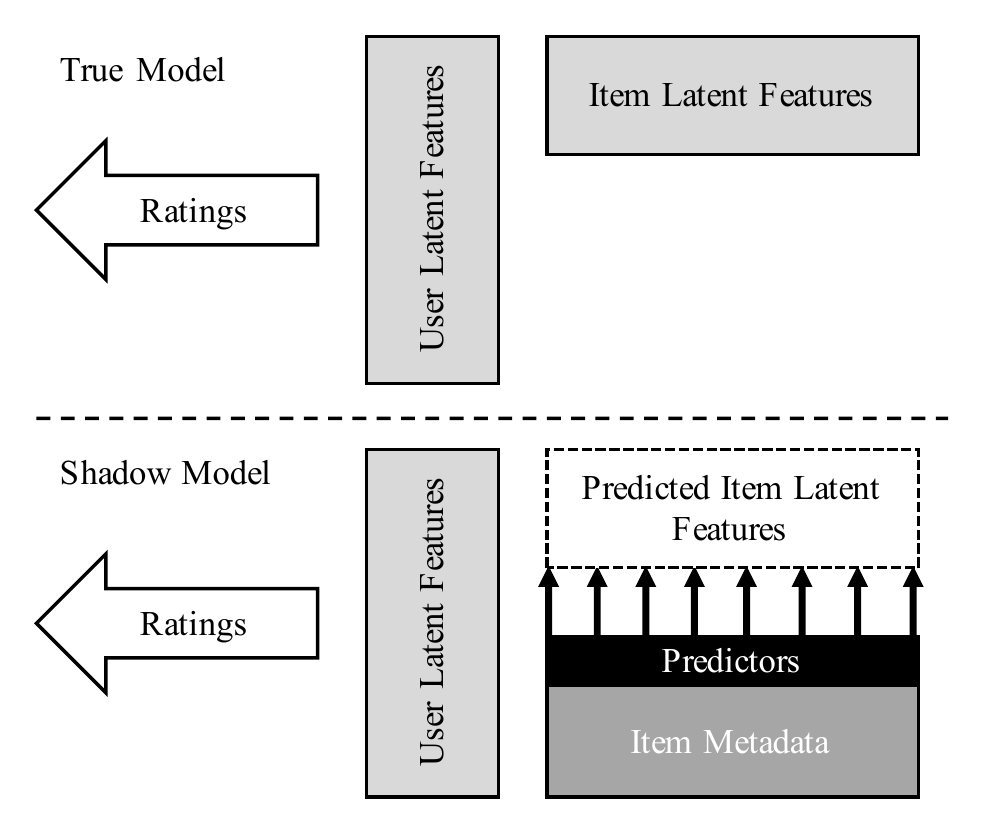}
  \caption{Shadow models constructed by training predictors for latent
    features from interpretable meta-data, and composing these
    predictors with the rest of the system.}
  \label{fig:shadow}\pxm{show agreements in figure}
\end{figure}

The first challenge is that learnt latent factors are constants
uninterpretable to humans; any explanations in terms of these factors
would be unintelligible.
In order to address this problem, we learn a mapping from
interpretable features to these latent factors.
We then compose the mapping with the rest of the model. 
In our setting, we compose the interpretation of item latent factors
with user latent factors to make recommendations (see
Fig.~\ref{fig:shadow}). 
We call the composed model, a \emph{shadow model}.

Our second challenge is that this composed shadow model still remains
too complex for direct interpretation.
However, since the shadow model expresses ratings in terms of
interpretable features, we can leverage existing model explanation
techniques~\cite{datta2016algorithmic,ribeiro2016lime}.
In particular, in this paper, we determine influential features using
an existing technique \cite{datta2016algorithmic} (see
Fig.~\ref{fig:results:qii} for an example).
Note that the purpose of the shadow model is not to supplant the
recommender system, but to interpret its predictions.

The third challenge is maintaining correspondence between
interpretations and the models they explain.
Re-expression of a system via a shadow model does not guarantee that
the interpretations constructed from the shadow represent the
functioning of the original. 
In our approach, we substitute predicted item latent factors but keep
the remaining structure of the recommender system intact. 
Therefore the effects of the item factors on recommendations in the
shadow model are identical to those of the original. 
Demonstrating a level of accuracy in predicting both the latent
factors, and the resulting recommendations, we can claim that our
interpretations are meaningful because \textbf{the shadow model makes
  similar recommendations for similar reasons.}

Results of user studies\cite{tintarev2007effective} indicate features
most important to movie recommendations largely include interpretable
features which we find can be derived from auxiliary sources, such as
average rating and keywords.
In our example LFI interpretation of a movie recommender system, we
predict the latent factors from such important interpretable features
and others derived from auxiliary data sources including IMDB and
DBTropes~\cite{data-imdb,data-dbtropes}.\pxm{todo: convert to url
  footnotes?}
An interpretation for a given recommendation thus indicates important,
human-understandable features behind it, e.g., a high recommendation
for Star Trek arose for a particular user because its genre is sci-fi,
and it has keywords indicating action in space.
Since the recommendations of the shadow model are close to the latent
model, this interpretation also serves as an interpretation of the
recommendations of the latent model.

As a proof of concept, we apply our techniques to a movie recommendation
system based on matrix factorization over the popular MovieLens
dataset with data integrated from several other movie databases,
producing interpretable explanations for recommendations.

This technique of training an approximate, but interpretable shadow
model for LFI is similar in spirit to approaches to explaining other
machine learning
systems~\cite{craven1995extracting,ribeiro2016lime,sanchez2015towards}.
An important difference is that prior work has explored this idea
using the features present in the task itself, or using pre-defined
mappings to an interpretable space. 
Instead, we use externally available interpretable features and
\emph{learn} the mapping to an interpretable space.
We also differ from existing approaches that attribute meaning to
latent factors, e.g.
with topic models\cite{rossetti2013towards}, in that the constructed
shadow model is itself a recommendation model, albeit with
interpretable inputs, and is therefore amenable to existing
explanation techniques for machine learning models.
We demonstrate this point by applying a recently developed input
influence measure\cite{datta2016algorithmic} to build interpretable
explanations for recommendations.
We focus on this approach because it does not make assumptions about
the complexity of the models involved and allows us to tailor
explanations to individual users (or individual recommendations).
User studies have identified the latter as the most important aspect
of effective explanations\cite{tintarev2007effective}.

This paper makes the following contributions:

\begin{itemize}
\item{} We present Latent Factor Interpretation (LFI), a method for
  interpreting models by leveraging expressions of latent factors
  in terms of human-understandable features, and develop it in the
  particular setting of collaborative filtering. 
\item{} We demonstrate how the approach applies to a real world
  use-case of a movie recommendation system trained from the MovieLens
  dataset and integrating auxiliary data from IMDB and DBTropes.
\item{} We demonstrate the accuracy of the approach for
  matrix-factorized models by constructing movie recommendation
  explanations for synthetic individuals with known preferences.
\end{itemize}

This paper is structured with a brief background
(\refsec{sec:background}) on matrix factorization for recommender
systems, %
and quantitative input influence which serve as the building blocks of
our approach or its evaluation later in this paper.
We then describe the construction of the shadow model and computation
of influence as means for interpreting recommendations
(\refsec{sec:methods}).
We demonstrate the utility of our approaches using synthetic and real
use-cases derived from the MovieLens\cite{data-movielens}\pxm{todo:
  replace with url-footnote?} 
movie database augmented with various information sources
(\refsec{sec:results}). 
We discuss related work (\refsec{sec:related}) and conclude with a
summary of contributions and directions for future work
(\refsec{sec:conclusions}).

\section{Background}\label{sec:background}

In a general sense recommender systems discover
liked items, such as movies, previously not encountered by users.
Numerous recommender systems have been proposed in literature, making
use of varying forms of data and providing a variety of types of
recommendations\cite{adomavicius2005toward}. 

In this section, we discuss a particular type of
collaborative recommender system based on matrix factorization
(\refsec{sec:background:matrix}). 
We conclude the section with an overview of \emph{quantitative input
  influence}, the main tool we will employ to construct explanations
for recommendations (\refsec{sec:background:qii}).

\subsection{Matrix factorization for
  Recommendations}\label{sec:background:matrix}

Recommendation systems, as the name implies, are models that give
recommendations to users regarding items they would enjoy or prefer.
Formally, we are given a set of $ n $ users a set of $ m $ items, and
a sparse $n$ by $m$ matrix of ground-truth ratings $ R $ and need to
fill in the missing elements of the matrix, that is, predict ratings.

\newcommand{\vect}[1]{\mathbf{#1}}

A state of the art method for constructing recommendation models is
via \emph{matrix factorization} \cite{koren2009matrix}.
The technique associates with each user a set of preferences over some
$ k $ number of \emph{latent features} and with each movie associates
a measure of expression of those $ k $ features.
Formally the model is composed of a $ u $ by $ k $ matrix $ U $ and a
$ i $ by $ k $ matrix $ I $ and the predicted rating for each user
movie pair is described by the $ u $ by $ i $ matrix
$ \hat{R} \defeq U I^\top $.
Thus each prediction for the rating of item $ i $ by user $ u $, or
$ \hat{t}_{ui} $ is the dot product of the $k$-length vector
$ \vect{u}_u $, expressing that user's preferences for the $ k $
latent factors, and the $k$-length vector $ \vect{i}_i $, expressing
the extent with which item $ i $ exhibits those latent factors.
The model \emph{factors} the ground truth matrix $ R $ into the matrix
product of $ U $ and $ I^\top $.
The choice of $k$, or \emph{rank}, varies.

Several algorithms exist for this task though this choice not
important in this paper.
Our experimental results are based on the implementation of
\emph{alternating least squares} in Apache Spark\pxm{todo: url
  footnote?}. 
Our implementation and experiments are available \ifdefined\arXiv
online\cite{artifact}\else in an anonymized
form\cite{artifact-anonymous}\fi.

\subsection{Quantitative input influence}\label{sec:background:qii}

We now briefly review a family of measures presented in Datta et
al.~\cite{datta2016algorithmic} called \emph{Quantitative Input
  Influence} (QII), that measures the influence of a feature on the
outcomes of a model.
QII can be tailored to a particular \emph{quantity of interest} about
the system, such as the outcomes of a model over a population, the
outcome for a particular instance or other statistics of the system.
We use this influence measure to identify influential metadata in
shadow models.
In particular we will use QII to measure the influence of metadata on
the predicted ratings for a specific user and movie pair.

At its core, QII measures the influence of features by breaking their
potential correlations with other input features. This focuses
measurements on the explicit use of a particular feature and not on
use via correlated other features.

Formally, given an a model $m$ that operates on a feature vector
$\vect x$, the influence of a feature $i$ is given by the expected
difference in outcomes when feature $i$ is randomly perturbed:

\begin{equation*}
\iota_{m, \vect x}(i) \defeq \mathbb{E}_{y_i}\left[  m(\vect x) -
      m(\vect x_{-i} y_i)    \right]
\end{equation*}

The expectation in the above quantity is over samples of the $i^\text{th}$ feature
$y_i$, which is drawn independently from its marginal distribution.

\section{Methods}\label{sec:methods}

Our approach to interpreting recommender systems based on matrix factorization
comprises of two steps. First, we use publicly available interpretable features 
(i.e., metadata) about items as interpretable features
to predict latent factors of these items. We then compose these models
for predicting latent factors into models that predict the outcomes
for particular users. Second, this shadow model composed of predictors for the latent
factors is used to generate human-understandable explanations of outcomes
by identifying the most influential interpretable features.

\subsection{Metadata sources}

\pxm{todo: remove section?} 
In case of movies, we use several sources of publicly available
metadata attributes such as genres, directors etc. 
that are one-hot encoded to obtain numerical features.

\subsection{Shadow Model}

We assume that we are given a matrix of interpretable attributes $A$, with
one row $\vect{a}_i$ for each item $i$. For each item latent factor $j$, we train
a predictor $f_j$  such that $f_j(\vect{a}_i) \approx \vect{i}_{ij}$.
Composing these predictors, the final predicted recommendation for a user $u$ and item $i$
can be approximated as follows:

\[\hat{r}_{ui} = \vect{u}_u\cdot\vect{i}_i = \sum_{j = 1}^k \vect{u}_{uj}\vect{i}_{ij} \approx
  \sum_{j = 1}^k \vect{u}_{uj}f_j(\vect{a}_i).
\]

Consequently, we use the composed model $\tilde{r}_u(\vect{a}) = \sum_{j = 1}^k
\vect{u}_{uj}f_j(\vect{a})$ as a model that predicts the outcomes of the system
for a movie with interpretable attributes $\vect{a}$ and user $u$. This shadow model
is more interpretable insofar as it maps interpretable attributes to
ratings. However, it is still fairly complex. Therefore, to interpret the
behavior of the shadow model $\tilde{r}_u$ on a point $\vect{a}$, we examine
the influences of interpretable attributes using QII.

\subsection{Interpreting the Shadow Model}

We interpret the shadow model by measuring the quantitative input
influence of all metadata features
on its output.
This can be measured either on the output of a particular user-item
pair, in which case the question being answered is ``why were you
given this recommendation?''
or the entirety of the model's predictions for this user over all
items, in which case the measure would be answering ``what has the
model inferred about your preferences in general?''.
In its raw form, an interpretation takes the form of a list of
feature-influence pairs but can be naturally visualized as in
Figure~\ref{fig:results:qii}.%

\subsection{Measuring latent factor accuracy}\label{sec:method:measuring}

We measure the quality of the shadow model by computing the mean
absolute error of its predictions compared to the original model,
that we call \emph{baseline}.

Another metric of the quality of the shadow model is how close it
agrees with the original model on the latent factors themselves.
For each factor, we compute the mean absolute error (MAE) of latent
factor prediction. 
Averaging over all factors, we get a measure of the overall latent
factor agreement.

\section{Results}\label{sec:results}

We evaluated our methods on movie recommendations after integrating
ratings data with several additional sources of movie metadata. 
We note some relevant details about the datasets we used in
\refsec{sec:results:datasets} and briefly describe our implementation
in \refsec{sec:results:implementation}.

In \refsec{sec:results:predict} we describe our experiments over
several combinations of parameters of both the recommender system
itself and the shadow model. 
We find that the overall performance improves with higher ranks, and
decision tree models perform the best in shadow models, although there
can be a trade-off between rating and latent factor agreement.

We present the recommendation interpretations we can derive using
these predictions in \refsec{sec:results:interpret}, finding that
usually only a relatively small number of metadata features is
influential in the final decision.
Finally, in \refsec{sec:results:signal} we describe some experiments
on synthetic data, with which we verify that this approach can derive
the true causes of recommendations.

\subsection{Datasets}\label{sec:results:datasets}

The source of our data was MovieLens 20M
Dataset\cite{data-movielens}\pxm{url-footnote?}, which contains
approximately 20 million ratings of 27,000 movies by 138,000 distinct
users.
Ratings are on a 1-5 integer $\bigstar$ range. 

Additionally we included various movie features from the Internet
Movie DataBase (IMDB)\cite{data-imdb}\pxm{url-footnote?} 
and DBTropes data\cite{data-dbtropes}\pxm{url-footnote?}, a
machine-readable snapshot of TV Tropes.

Overall, the three sources of data contain a wealth of movie
information.
Of the most relevant factors for recommendations as noted in user
studies\cite{tintarev2007effective}, a substantial portion can be
determined to some extent from the metadata we have collected.

\paragraph{Pre-processing}

We used all movie ratings from MovieLens 20M dataset for constructing
a recommender system. 
For subsequent steps, however, we performed several pre-processing
steps.

We encode nominal features via one-hot encoding, and in a feature
selection step, we dropped those not meeting a minimal entropy
threshold.
For training and evaluating explanations, we also pruned away movies
with missing or negligible metadata. 
We justify this step as a deployed recommendation explanation system
could itself recognize its lack of metadata and notify users of said
fact instead of providing a potentially inaccurate explanation for its
recommendation.

\subsection{Implementation}\label{sec:results:implementation}

Our implementation is based on a set of Python programs that make use
of the Apache Spark library\pxm{url footnote?} 
for model training and evaluation.

\subsection{Learning recommendations and latent
  features}\label{sec:results:predict}

The MovieLens ratings constituted the sparse user-item input matrix
for the training of a recommender system.
This data is also the \emph{ground truth} for evaluation purposes
later in this section.
The ground truth was processed with alternating least squares matrix
factorization algorithm, as implemented in Apache Spark MLlib, which
outputs two matrices: \emph{user features} and \emph{movie features},
which encode user preferences and movie properties along
low-dimensional space of \emph{latent features}.

\begin{figure}
\includegraphics[width=3.3in]{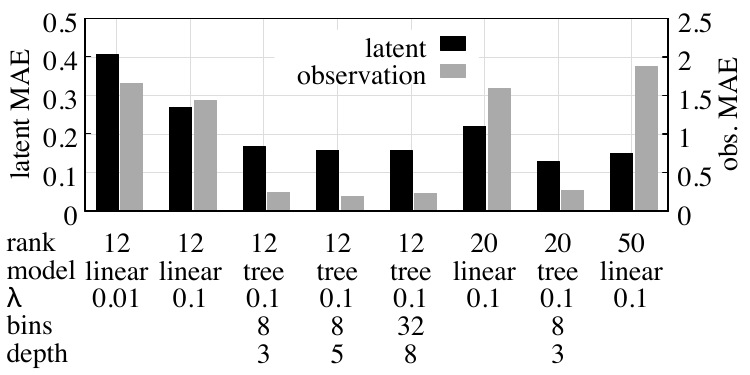}
\caption{\label{table:results:parameters} Mean absolute error of shadow
	models compared to real models with different parameters. Two metrics
	are shown: error in predicting latent factors of the real model, and error
	in predicting the ratings the real model gives}
\end{figure}

We trained recommender systems, constructed shadow models, and
measured the prediction error of both individual
latent factors (which can be then averaged across all of them)
and the overall predicted ratings, iterating over
several possible parameters (rank and regularization parameter for the
recommender, type of the shadow model (linear or decision tree), and
the number of bins and maximal depths for tree models). The results of
these experiments are summarized in Figure
\ref{table:results:parameters}.

It can be seen that linear models consistently perform worse on both
metrics than decision trees. 
However, the difference in performance is much higher on observational
agreement than on latent factor agreement. 
We hypothesize that it could be due to to the linear regression models
having a consistent bias that adds up during the matrix multiplication
which is consistent with our
observations.

The observational agreement for linear models also gets worse with higher ranks,
whereas the latent factor agreement gets better, which is also consistent with the
bias hypothesis.

In our experiments, one model (Rank 50, tree, lambda 0.1, depth 5, bins 32) performed
best on both metrics, but in general, there can be a trade-off between them. Namely,
if we exclude the best-performing model, we can see that different models are the
second-best for latent factor agreement (rank 20, tree, lambda 0.1, 8 bins, depth 3)
and observational agreement (Rank 12, tree, lambda 0.1, 8 bins, depth 5),
although their performance is reasonably close to each other.

\subsection{Interpreting recommendations}\label{sec:results:interpret}

\begin{figure*}[t!]
  \centering
  \begin{subfigure}[t]{0.45\textwidth}
    \includegraphics[width=3in]{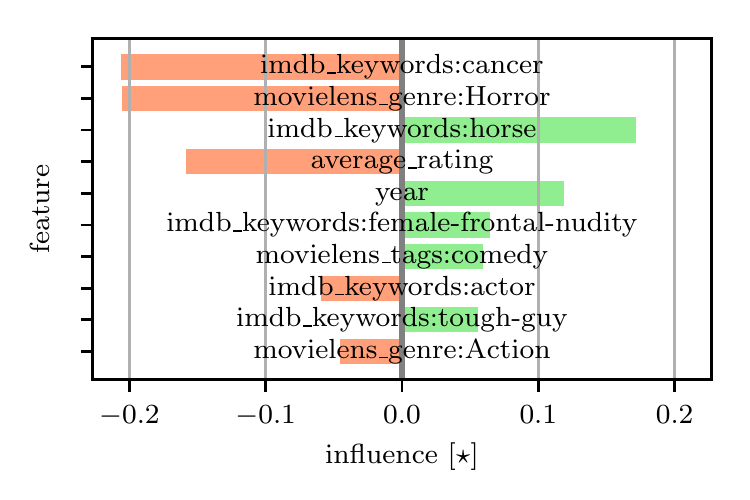}
    \caption{User 7's recommendation about Lake Placid (1999)}
  \end{subfigure}~
  \begin{subfigure}[t]{0.45\textwidth}
    \includegraphics[width=3in]{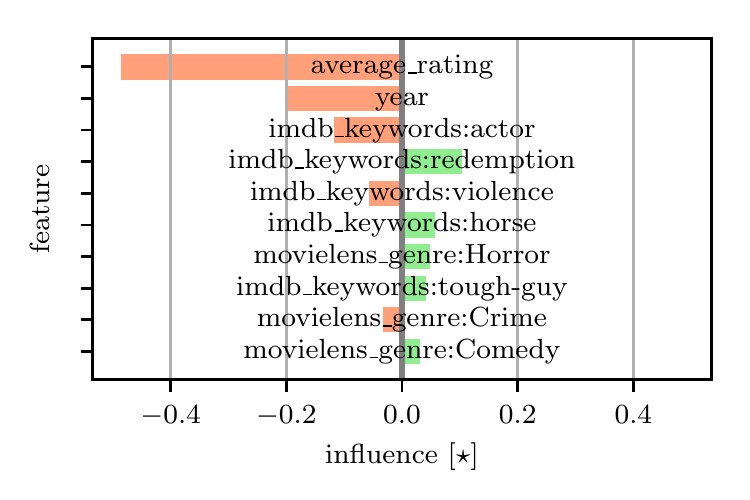}
    \caption{User 21's recommendation about Lake Placid (1999)}
  \end{subfigure}

  \begin{subfigure}[t]{0.45\textwidth}
    \includegraphics[width=3in]{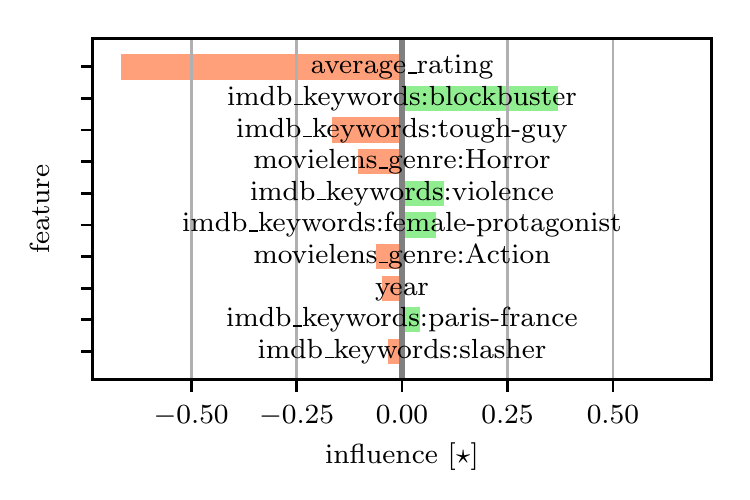}
    \caption{User 7's recommendation about Inspector Gadget (1999)}
  \end{subfigure} ~
  \begin{subfigure}[t]{0.45\textwidth}
    \includegraphics[width=3in]{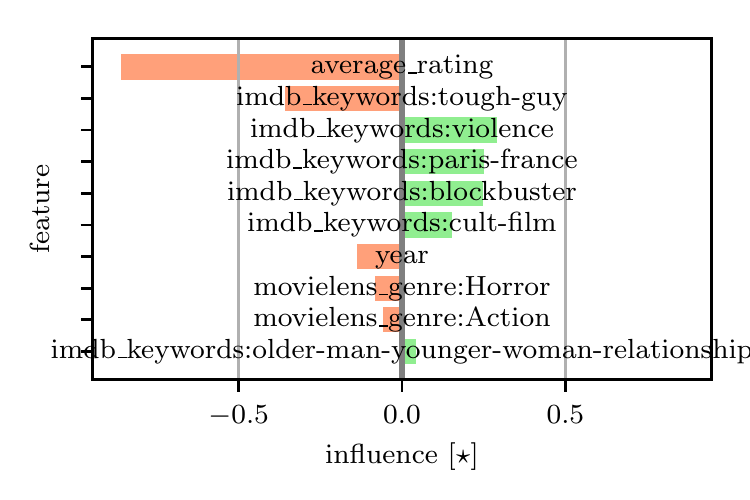}
    \caption{User 21's recommendation about Inspector Gadget (1999)}
  \end{subfigure}

  \caption{\label{fig:results:qii}A sampling of QII-based
    recommendation interpretations based on shadow models over the
    MovieLens 20M dataset for User 7 (left) and User 21 (right). 
    \pxm{what were the actual recommended ratings for these movies and
	these users?}\sophia{unknown, and can't easily retrieve}
  }
\end{figure*}

To construct interpretations of the recommendations produced by the
shadow model, we measure the influence of each of the metadata
features on the rating the shadow model produces. 
For a particular recommendation (user,movie pair), the definition of
influence of a metadata feature (see \refsec{sec:background:qii}) in
this setting measures the expected change in the output of the
recommender (the rating) if we substitute a fresh value for only that
metadata feature with one sampled independently from its marginal,
while all the other metadata features are kept fixed.

Several examples of the resulting influence measures can be seen in
Figure~\ref{fig:results:qii}. 
For two users, we see the top 10 most influential features in their
recommendation for two different movies.
We order the influential metadata features on the y-axis and chart
their influence (which can be measured in $\bigstar$) on the x-axis. 

\subsection{Validation against Known
  Preferences}\label{sec:results:signal}

In the absence of user studies, we simulated users generating movie
ratings based on known rules in order to verify the hypothesis that
our system can detect the true user preferences. 
For this approach we generated a synthetic dataset based on a simple
user preference and rating simulation.

In the simulation, a set of user preferences were generated to be
expressible exactly in terms of randomly selected movie metadata
features.
Randomly assigned to a user, these preferences either increased or
decreased their simulated movie ratings. 
Ratings were generated for randomly selected set of movies for each
user.

We trained a matrix factorization model on the synthetic data set, and
performed our analysis based on the shadow model construction and QII
measurement.
We then calculated a score from 0 to 1, indicating the closeness of
the measured QII values compared to the known metadata features that
were actually used to simulate ratings.
This score is then compared to its value calculated with respect to
sets of independently randomized or partially randomized user
preferences (not the ones used to generate the synthetic dataset). 
This comparison is done in expectation that the real evaluation score
will perform statistically significantly better on the synthetic
personalities than on only partially related semi-random
personalities, and in turn better yet than on fully random
personalities.\pxm{I adjusted the previous to fix grammar but I might
  have also changed the meaning, is it correct?}
As demonstrated in Table~\ref{table:ttest}, this hypothesis holds
given sufficiently large sample sizes.

Additionally, we manually constructing a matrix factorization model
that directly encodes simulated user preferences, and found that such
a system perfectly captures synthetic personalities rather than just
better than controls.
This suggests that some of the errors in determining user preferences
could be due to the recommender system trainer's inability to learn
the right predictive model (even if one exists) rather than due to
inaccuracies in our system of shadow models.
\begin{table}
  \small
  \begin{tabular}{|p{1.3cm}|p{0.50cm}|p{0.50cm}|p{0.50cm}|p{0.60cm}|p{0.35cm}|p{0.75cm}|p{0.5cm}|} \hline
    \multirow{2}{*}{Parameters} & \multirow{2}{0.55cm}{t. mean} & \multirow{2}{0.55cm}{s.r. mean} & \multirow{2}{0.55cm}{r. mean} & \multicolumn{2}{|c|}{\mbox{t. > s.r.}} & \multicolumn{2}{|c|}{s.r. > r.} \\
    \cline{5-8}                 &      &      &      & p     & e.s. & p       & e.s. \\ \hline
    {\scriptsize N=20, 3 pr, rn 3, 15 h.e.f.} & 0.75 & 0.51 & 0    & {\scriptsize 6e-11} & 3.3  & 1e-20   & 14.2 \\ \hline
    {\scriptsize N=20, 8 pr, rn 3, 40 h.e.f.} & 0.26 & 0.2  & 0    & 0.03  & 0.8  & 8e-11   & 4.2 \\ \hline
    {\scriptsize N=20, 8 pr, rn 8, 40 h.e.f.} & 0.4  & 0.3  & {\scriptsize 2e-4} & 0.02  & 0.5  & 1e-23   & 4.5 \\ \hline
    {\scriptsize N=20, 10 pr, rn 15, any 250} & 0.22 & 0.19 & 0    & 0.1   & 0.5  & 7e-11   & 4.2 \\ \hline
    {\scriptsize N=49, 10 pr, rn 15, any 250} & 0.22 & 0.19 & 0    & 0.02  & 0.5  & {\scriptsize 1.5e-27} & 4.6 \\ \hline
  \end{tabular}%
  \caption{\label{table:ttest} Synthetic data set hypotheses testing.
    The parameters of the experiments include: sample size, number of user preference profiles, rank of the matrix factorization model, and the strategy of selecting features for generating profiles.
    ``h.e.f.''
    stands for highest-entropy features, ``any 250'' stands for any features with more than 250 non-zero values, ``s.r.''
    stands for semi-random, ``r.''
    stands for random, ``t.''
    stands for true, ``e.s.''
    stands for effect size, ``pr'' stands for profiles, ``rn'' stands for rank.}
\end{table}

\section{Related Work}\label{sec:related}

Existing approaches to address the interpretability of latent factors
either attempt to associate them with some item content, or to present
them via the relationships they encode in the items and users of a
system. 
We summarize these approaches in this section.
We also discuss the relationship of our methods to other approaches
for making machine learning interpretable via shadow models and
interpretability constraints.

\paragraph*{Associations}

Rossetti et al.\cite{rossetti2013towards} use topic modeling to
extract topics from movie descriptions and then associate topics to
latent factors in a matrix factorized model.
Their topics are of the form of bags of words and are not as directly
interpretable as movie features we consider in our work.
Further, they develop said association so that the recommendation
model can be portable; new users specify their preferences on topics
and the technique can then provide them recommendations by injecting
the topic-latent matrix within the usual matrix factor model.

\paragraph*{Presentation}

Koren et al.\cite{koren2009matrix} show that movies and users can
sometimes be understood in terms of their proximity to other movies
and users.
Plotting users and movies according to their latent features or
certain projections can result in recognizable clusters.
These clusters can then be suggestive of user personalities and of
movie characteristics that may have not been part of their extrinsic
characteristics.
For example, they show how groups of movies form clusters that roughly
correspond to movies with strong female leads and fraternity humor.

In a related line of work, Hernando et al.\cite{hernando2013trees}
present a design of a tool in which recommendation explanations are of
a form of a graph with users and movies as nodes, arranged to
designate proximity in the latent feature space.

\paragraph*{Shadow Models}

Our approach of training an interpretable shadow model that mimics the
behavior of the true uninterpretable model is similar to a general
approach for explaining machine learning
algorithms~\cite{thrun1994ann,craven1995extracting,lehmann2010ann,ribeiro2016lime}, and has also been applied to matrix factorization
techniques~\cite{sanchez2015towards}.
These approaches either use features present in the input space or map
to an interpretable space using handcrafted mappings. 
LFI uses externally provided interpretable features and \emph{learns}
a mapping to the latent space.
Similar to us, Gantner et al.\cite{gantner2010learning} use externally
provided interpretable features in order to train a shadow model.
They do this to alleviate the cold-start problem as their shadow
models allow them to recommend items without ratings.
Our focus, however, are explanations for recommendations.
Whereas they can recommend rating-less items, we can provide an
explanation for such recommendations.
We theorize that explanations can further alleviate the cold start
problem, as explanations for recommendations of new items can
encourage users to rate them.

\paragraph*{Interpretability Constraints}
An orthogonal approach to adding interpretability to machine learning
is to constrain the choice of models to those that are interpretable
by design. 
This can either proceed through regularization techniques such as
Lasso~\cite{Lasso} that attempt to pick a small subset of the most
important features, or by using models that structurally match human
reasoning such as Bayesian Rule Lists~\cite{letham2015}, Supersparse
Linear Integer Models~\cite{ustun13}, or Probabilistic
Scaling~\cite{ruping2006}. 
For recommender systems, one approach that belongs to this family is
non-negative matrix factorization (NMF) (see \cite{lee99}), that
enforces a level of interpretability by constraining latent features
to be positive. 
Even for NMF, the mapping to interpretable features could be useful
for discovering the concepts encoded in these latent factors.

\paragraph*{Cold-start in collaborative
  filtering}%

Collaborative recommender systems like those based on matrix
factorization suffer from \emph{cold-start} problem:
recommendations cannot be provided for new users or new items without
an existing set of ratings by those users or for those items,
respectively.

Several works address this problem by establishing connections between
latent factors and content features, as we do in our approach for
constructing explanations\cite{gantner2010learning}. However, our evaluation
metrics are optimized for a different goal.

\section{Conclusions and Future Work}\label{sec:conclusions}

We describe a method for interpreting recommendations of latent factor
models for collaborative filtering. 
We construct a shadow model that agrees with the latent factor model
in its predictions and its latent factors themselves, which are
predicted from interpretable features available from auxiliary data
sources. 
The metadata-expressed latent factors are then used to make
recommendations like in the original model. 
In contrast to prior work, the shadow model is not interpretable by
design. 
In fact, it is more complex than the original model. 
However, since its input features are interpretable, its
recommendations can be explained using input influence measures from
prior work. 

We apply this method to a movie recommendation system based on matrix
factorization over the popular MovieLens dataset with auxiliary data
from IMDB and TV Tropes, producing interpretable explanations for
recommendations. 
We find that the influence measures that quantify the impact of
interpretable features on recommendation ratings in the shadow model
are a reasonable and concise way of interpreting the functioning of the
latent factor recommender system.

There are several avenues for future work. 
One interesting direction is the design of explanations for hybrid
content/collaborative recommender systems which use some interpretable
features along with user ratings, making them amiable to influence
measures, though only partially via their interpretable inputs.
Other open questions include formally characterizing conditions under
which this interpretation method effectively reveals user preferences
as well limits that arise from lack of informativeness in auxiliary
data sources. %
A related direction involves validating these explanations with real
users through user studies.

\ifdefined\anon
\else
\paragraph*{Acknowledgments}
\begin{small}
This work was developed with the support of NSF grant CNS-1704845 as
well as by DARPA and the Air Force Research Laboratory under agreement
number FA8750-15-2-0277. 
The U.S. 
Government is authorized to reproduce and distribute reprints for
Governmental purposes not withstanding any copyright notation thereon. 
The views, opinions, and/or findings expressed are those of the
author(s) and should not be interpreted as representing the official
views or policies of DARPA, the Air Force Research Laboratory, the
National Science Foundation, or the U.S. 
Government.
\end{small}
\fi

\bibliographystyle{named}
\bibliography{recommend}

\end{document}